\documentclass[12pt]{article}

\input{epsf}
\usepackage[dvips]{graphicx}
\usepackage{cite}
\def\1ad{\mbox{\normalsize $^1$}}
\def\2ad{\mbox{\normalsize $^2$}}
\def\3ad{\mbox{\normalsize $^3$}}
\def\4ad{\mbox{\normalsize $^4$}}
\def\5ad{\mbox{\normalsize $^5$}}
\def\6ad{\mbox{\normalsize $^6$}}
\def\7ad{\mbox{\normalsize $^7$}}
\def\8ad{\mbox{\normalsize $^8$}}

\def\dj{\hbox{d\kern-0.347em \vrule width 0.3em height 1.252ex depth
-1.21ex \kern 0.051em}}

\newcommand{\be}{\begin{equation}}
\newcommand{\ee}{\end{equation}}
\newcommand{\ben}{\begin{equation*}}
\newcommand{\een}{\end{equation*}}
\newcommand{\ba}{\begin{eqnarray}}
\newcommand{\ea}{\end{eqnarray}}
\newcommand{\ban}{\begin{eqnarray*}}
\newcommand{\ean}{\end{eqnarray*}}
\newcommand{\brr}{\begin{array}}
\newcommand{\err}{\end{array}}
\newcommand{\bc}{\begin{center}}
\newcommand{\ec}{\end{center}}

\newcommand{\bea}{\begin{eqnarray}}
\newcommand{\eea}{\end{eqnarray}}
\newcommand{\bean}{\begin{eqnarray*}}
\newcommand{\eean}{\end{eqnarray*}}

\newcommand{\ie}{\mbox{\it i.e.~}}

\newcommand\lsim{\mathrel{\rlap{\lower4pt\hbox{\hskip1pt$\sim$}}
    \raise1pt\hbox{$<$}}}
\newcommand\gsim{\mathrel{\rlap{\lower4pt\hbox{\hskip1pt$\sim$}}
    \raise1pt\hbox{$>$}}}
\begin{document} 

\begin{titlepage}

\title{\bf {A Way to Reopen the Window \\ for Electroweak Baryogenesis}}

\author{ G. Servant }

\maketitle
\begin{center}
{\it Enrico Fermi Institute, University of Chicago, Chicago, 
IL 60637\\
High Energy Physics Division, Argonne National Laboratory, Argonne, IL 60439.}
\end{center}
\vskip 3pt

\vglue .5truecm

\begin{abstract}
\vskip 3pt

\noindent
We reanalyse the sphaleron bound of electroweak baryogenesis when allowing deviations 
to the Friedmann equation. These modifications are well motivated in the context 
of brane cosmology where they appear without being in conflict with major 
experimental constraints on four-dimensional gravity.
While suppressed at the time of nucleosynthesis, these corrections 
can dominate at the time of the electroweak phase transition and in certain cases provide
 the amount of expansion needed to freeze out the baryon asymmetry without requiring 
a strongly first order phase transition. The sphaleron bound 
is substantially weakened and can even disappear so that the 
constraints on the higgs and stop masses do not apply anymore.
Such modification of cosmology at early times therefore reopens the parameter 
space allowing electroweak baryogenesis
which had been reduced substantially given the new bound on the higgs
 mass imposed by LEP. 
In contrast with previous attempts to turn around the sphaleron bound using alternative 
cosmologies, we are still 
considering that the electroweak phase transition takes place in a radiation dominated
 universe. The universe is expanding fast because of the modification of the Friedmann 
equation itself without the need for a scalar field and therefore evading the problem 
of the decay of this scalar field after the completion of the phase transition and the 
risk that its release of entropy dilutes the baryon asymmetry produced at the transition.

\end{abstract}

\vskip 26pt

\begin{flushright}
ANL-HEP-PR-01-120\\
EFI-01-55
\end{flushright}

\end{titlepage}

\section{Introduction}

Since the appearance of the article by Kuzmin, Rubakov and Shaposhnikov in 1985 
\cite{Kuzmin:1985mm} who suggested that baryon number violation at high
 temperature in the Standard Model might have played a role in
 the early universe and set the basis for the theory of electroweak baryogenesis,
there have been extensive studies which explored 
with great scrutiny the details of the mechanism and its viabiblity both in
 the Standard Model and in the Minimal Supersymmetric Standard Model. The fact 
that the creation of the baryon asymmetry of the universe is
 in principle possible in the standard electroweak theory is indeed very appealing 
since it relies only on physics accessible at present colliders in contrast with GUT 
or Affleck--Dine scenarios. Electroweak physics is not though the only ingredient.
Electroweak baryogenesis makes use of an intricate interplay between particle 
physics and cosmology which is much more involved than in the theory of nucleosynthesis.
Let us summarize very briefly the main idea: In a universe with conserved vanishing $B-L$
 charge and standard Friedmann--Robertson--Walker (FRW) cosmology, the baryon
 asymmetry of the universe (BAU) can only be produced at the electroweak phase transition (EWPT).
 To fulfill Sakharov's out-of-equilibrium requirement for baryogenesis \cite{Sakharov:dj}
the phase transition 
has to be first order. In that case, bubbles of non zero higgs field {\it vev} nucleate 
from the symmetric vacuum 
and as they expand, particles in the plasma interact with the moving phase 
interface (bubble wall) in a CP violating
way producing a CP asymmetric flux of particles into the symmetric phase. This 
CP asymmetry is converted into a baryon asymmetry by baryon number violating sphaleron 
processes in front of the wall. As the phase interface propagates, the baryons 
enter into the broken phase where baryon violation is inefficient and therefore baryon 
asymmetry frozen out \cite{Cohen:py}.

To compute the resulting net baryon asymmetry produced during the electroweak phase 
transition one 
needs to know: (1) The details of the non trivial baryon number violating processes 
at high temperature 
(B violation is achieved
 thanks to sphaleron transitions interpolating between two neighboring N-vacua of the 
 electroweak gauge theory). (2) The amount of departure from equilibrium at the electroweak
 phase transition. This necessitates an accurate calculation of the higgs effective 
potential at high temperature. (3) The generation and propagation of CP-violating particle fluxes.
(1) is now well under control. The rate of B-violating sphaleron processes at high temperature
 has been performed accurately and checked independently using analytic perturbative and non
 perturbative techniques both in the broken 
phase \cite{Carson:1990jm,Baacke:1993aj,Moore:1998ge} and
in the symmetric phase \cite{Moore:1998mh,Bodeker:1999zt}. (2) has also been quite well understood. It 
is at this point that cosmology enters into the game when comparing B violation rate with the 
expansion rate of the universe. One of the strongest constraint in electroweak baryogenesis 
comes from the 
requirement that 
baryons produced at the bubble wall are not washed out by sphaleron processes after they 
enter the broken phase. Imposing that sphaleron processes are sufficiently suppressed in
 the broken phase at 
the critical temperature leads to the so-called {\it sphaleron bound}: 
$\langle \phi(T_c) \rangle /T_c \gsim 1$. $\langle \phi(T_c)\rangle$ is the order parameter 
of the phase transition, the {\it vev} of the higgs field at the critical temperature. The derivation 
of this bound uses the expression of the expansion rate of the universe $H$ given by the 
Friedmann equation evaluated in a radiation dominated universe. 
$\langle \phi(T_c) \rangle /T_c $ on the other hand is given by the analysis of the effective 
scalar potential at high temperature and depends sensitively on the higgs mass so that the
 sphaleron bound translates into a constraint on the higgs mass. 
It has been shown that in the Standard Model the phase transition is not first order but just
 a cross over for a higgs mass above $\approx 72$ GeV \cite{Kajantie:1996mn} so
 that electroweak baryogenesis is excluded given the present experimental lower limit of the 
Standard Model higgs mass. Another 
reason in favour of
 that conclusion is related to our third point (3): CP violation in the Standard Model
 is not enough to account for the baryon asymmetry of the universe\cite{Gavela:1994yf}. Consequently,
 efforts have concentrated on the 
Minimal Supersymmetric Standard Model which provides not only the possibility of achieving a first 
order phase transition but also 
additional sources of CP violation\cite{Myint:wi}. While
 there is a common agreement on the identification of the major
CP violating sources--the complex parameters $\mu$ and $M_2$ in the mass matrix of the charginos--
the result of the calculation of CP violating currents on the other hand remains an open question
\cite{Carena:2000id,Huber:2001xf,Cline:2000nw,Cline:2001rk}.
We will not consider this issue here and assume that there are sufficient sources of CP violation 
in the MSSM to account for the baryon asymmetry of the universe so that the strongest constraint 
on electroweak baryogenesis essentially comes from the sphaleron bound.

We will therefore focus now on point (2) and on the condition  
$\langle \phi(T_c) \rangle /T_c \gsim 1$. 
$\langle \phi(T_c) \rangle /T_c$ measures the strength  of the phase transition and
depends crucially on the presence of a light stop $\tilde{t}_R $
(with small left-right mass mixing parameter $\tilde{A}_t$). It is difficult to compute since 
it is radiatively
 induced and perturbation theory is expected to break down at high temperature. Nevertheless, 
the validity of perturbation theory and results at two loop 
level \cite{Espinosa:1996qw,Carena:1997ki} have
 been confirmed by
 lattice calculation \cite{Csikor:2000sq} which enhance $\langle \phi(T_c) \rangle /T_c$  by $10-15 \%$ 
compared with perturbative calculation. From the last results \cite{Quiros:2000wk} one can infer 
 the constraints imposed by the sphaleron bound:
\be
\label{window}
110 \; \mbox{GeV} \lsim m_h \lsim 115 \; \mbox{GeV} \; \; \mbox{and} \; 
\; 105 \; \mbox{GeV}\lsim m_{{\tilde{t}}_R} \lsim 165 \; \mbox{GeV}
\ee
As the LEP bound on the higgs mass has been pushed up to $m_h \gsim 115$ GeV, the window 
(\ref{window}) for electroweak baryogenesis has been seriously reduced.
A large value for the higgs mass requires in particular
a large stop mixing parameter $\tilde{A}_t$. On the other hand a large 
$\tilde{A}_t$ makes the phase transition weaker and leads to values of 
$\langle \phi(T_c) \rangle /T_c $ in conflict with the condition of preservation of the baryon 
asymmetry. 

In summary, given the last bound on the higgs mass, electroweak baryogenesis in the MSSM 
starts to be in 
bad condition. However, if it turns out that the window (\ref{window}) is ruled out by 
experiment shall we definitely claim the death of electroweak baryogenesis? Adding new 
particle physics ingredients to enhance the phase transition would start to be arbitrary and 
electroweak baryogenesis would lose its main advantage over other baryogenesis mechanisms
that it relies on a well motivated and well defined particle physics framework. If there is no
 reasonable way out on the particle physics side and if one still wants to believe that 
the baryon asymmetry of the universe was created at the electroweak phase transition the only way
to turn around the sphaleron bound is to modify cosmology and question the common assumption 
made on the thermal history of the universe. There have been only a few attempts 
in the past to study the possibility of relaxing the sphaleron bound using alternative 
cosmologies\cite{Joyce:1996cp,Joyce:1997fc,Davidson:2000dw}. An interesting discussion has 
been offered in \cite{Joyce:1997fc}. 
All\footnote{For exhaustiveness, let us mention another type of alternative: If the EWPT 
was preceded by a color and charge breaking phase, the ratio $\langle \phi(T_c) \rangle /T_c $
could be much larger than in the usual scenario. However, it has been shown that such 
two-stage EWPT is not 
cosmologically viable \cite{Cline:1999wi}.} these works have in
 common the fact that they modify the usual assumption of a radiation dominated universe
 by assuming the existence of a scalar field driving the fast expansion of the universe
 at the EWPT. One drawback with this approach is that the decay of this scalar field 
after the completion of the phase transition may release entropy and dilute the baryon 
asymmetry produced at the transition. 

In this paper we take 
a different point of view and discuss a new cosmological 
solution to weaken the sphaleron bound. We still assume radiation domination at
 the EWPT. Instead, what we modify is the Friedmann equation itself in a way 
motivated by results in brane cosmology. While the usual expansion rate is recovered
 at late times in particular at nucleosynthesis, the expansion can be accelerated 
at high temperature with major consequences for electroweak baryogenesis as already
 suggested in \cite{Cline:1999ts}. In section II we present the derivation of the 
sphaleron bound and discuss how modifications of common assumptions on the cosmological evolution
 could substantially weaken that bound. In section III we motivate our parametrization
of deviations to Friedmann equation by reviewing generic aspects of brane cosmology and discuss 
under which general conditions these deviations may play a significant role at the time of the EWPT.
 Section IV makes some comments on the cosmology with a non standard Friedmann equation.
In section V we exhibit our numerical results describing the parameter space allowing 
the preservation of the baryon asymmetry.

\section{Derivation of the sphaleron bound}
 
 To derive the sphaleron bound, one starts with the``master equation'' of electroweak
 baryogenesis which gives  
the rate at which baryon number relaxes to its equilibrium value
 in the presence of fermions in the hot plasma \cite{Bochkarev:1987wf}:
\be
\frac{\partial n_B^{com}}{\partial t}=-V_B(t) n_B^{com} 
\label{master}
\ee
\be
V_B(t)={\cal M} \frac{\Gamma_s}{T^3}
\ee
$n_B^{com}$ is the density of baryons per comoving volume in an expanding universe,
$n_B^{com}=a(t)^3n_B$, where $a(t)$ is the scale factor of the universe. 
$V_B(t)$
is the rate of the baryon number non conserving processes. ${\cal M}$ is related to 
the number of fermionic degrees of freedom at thermal equilibrium which 
bias the free energy of the $SU(2)$ vacuum. In the Standard Model, 
${\cal M}=13N_f /2$, if one assumes that all fermionic degrees of freedom (including 
$e_R$) are at thermal 
equilibrium. In the MSSM, ${\cal M}=15N_f/2$ \cite{Cline:2000nw,Carena:2000id}.
The sphaleron rate, $\Gamma_s$,
 is the rate per
 unit time and unit volume of fluctuations with changing of the topological number. 
In the broken phase, a perturbative analysis
 gives \cite{Carson:1990jm,Arnold:1987mh,Khlebnikov:sr,Carson:1989rf}
\be
\label{gammas}
\Gamma_s \sim \eta \  T^4 \ {\zeta^7} \ e^{-\zeta}
\ee
where 
\be
\label{sph_rate}
\zeta(T)=E_{sp}(T)/T \ \ \ \mbox{and} \ \ \ 
\eta=2.8\times 10^5 \left(\frac{\alpha_w}{4\pi}\right)^4\kappa B^{-7}
\ee
$E_{sp}(T)=\sqrt{\frac{4\pi}{\alpha_w}}B{\phi(T_c)}$ is the energy of the sphaleron
 configuration \cite{Dashen:1974ck,Manton:1983nd}, 
in other words the energy 
barrier between two adjacent topologically non equivalent vacua of the SU(2) broken phase.
$B(\lambda/g^2)$ is evaluated numerically and is a monotonic slowly varying function of the higgs mass:
In the Standard Model, $B(0)=1.5 \rightarrow B(\infty)=2.7$ \cite{Manton:1983nd,Kunz:1992uh}. The 
value of $\kappa$ has 
originally been assumed to lie in the range 
$10^{-4}\lsim \kappa \lsim 10^{-1}$ \cite{Carson:1990jm}, however, 
recent non perturbative calculations \cite{Moore:1998ge} have shown that it is reasonable to use
 the perturbative estimate ({\ref{gammas}) provided that $\kappa \sim{\cal O}(1)$.

Integrating (\ref{master}) between the 
time $t_c$ at which the electroweak phase transition takes place and some time $t$ leads to
\be
\frac{n_B^{com}(t)}{n_B^{com}(t_c)}=e^{-\int_{t_c}^{t}V_B(t')dt'}\equiv {\cal S}
\label{integration}
\ee
where we have introduced the dilution factor ${\cal S}$. The most efficient dilution of the baryon 
asymmetry takes place just after the phase 
transition at a temperature near the critical temperature $T_c$ so that it is not a bad 
approximation (we will check it shortly) to replace the integral in (\ref{integration}) 
by the value of the integrand at $T=T_c$:
\be
\frac{n_B^{com}(t_c)}{n_B^{com}(0)}=e^{-V_B(t_c)\Delta t}
\ee
where $\Delta t$ is the time interval corresponding to a range of temperatures
 $\Delta T\sim T_c-T_2=(130-110)$ GeV, $T_2$ being the temperature at which dilution is inoperative.
We note that 
\be
n_B^{com}=a^3n_B=s^{com}\frac{n_B}{s}
\ee
where $s^{com}=a^3 s$ is the entropy per comoving volume. In a radiation
 dominated universe $s\propto g_*(T)T^3$. At this stage, we need the relation 
between time and temperature and that is where cosmology comes in. However, we want 
to remain as general as possible and will not assume any specific form of the Friedmann equation
for the moment. We will only assume 
that the universe is expanding adiabatically,
$d(\rho a^3)=-pd(a^3)$ ($p$ is the pressure), and that the universe is radiation 
dominated, leading to the relationship between the
 energy density $\rho$ and 
 the scale factor of the universe 
$a(t)$: $\rho(t)=\rho_0(a(t)/a_0)^{-4}$
where $\rho$ is given by
\be
\rho=\frac{\pi^2}{30}g_*T^4
\label{radiation}
\ee
$g_*$ counts the number of degrees of freedom at thermal equilibrium. At $T\sim100$ GeV, 
$g_*=106.75$ in the Standard Model. From (\ref{radiation}) we get
$a(t)\propto\rho(T)^{-1/4}\propto g_*^{-1/4}(T) T^{-1}$ so that $s^{com}$ scales as 
$s^{com} \propto g_*^{1/4}(T)$. We can rewrite the dilution as 
\be
{\cal S}=\frac{\frac{n_B}{s}(t)}{\frac{n_B}{s}(t_c)}\left(\frac{g_*(T)}{g_*(T_C)}\right)^{1/4}
\ee
${n_B}/{s}$ measures the baryon asymmetry. The value of the asymmetry today deduced 
from nucleosynthesis is \cite{Olive:1999ij} $1.7 \ 10^{-11} \lsim {n_B}/{s} \lsim 8.9 \ 10^{-11}$ so that
\be
{\cal S} \sim \frac{10^{-11}}{\frac{n_B}{s}(t_c)}
\ee
The value of the baryon asymmetry which can be produced at the EWPT, ${\frac{n_B}{s}(t_c)}$,
is still an open question \cite{Carena:2000id,Huber:2001xf,Cline:2000nw,Cline:2001rk}, however, it is 
generically difficult to produce a large
 asymmetry and one can reasonably write the bound ${{n_B}(t_c)/{s}(t_c)}\lsim 10^{-6}$ which means that 
for the baryon asymmetry not to be washed out, one demands ${\cal S}\gsim 10^{-5}$. This leads to 
\be
\ln\left(\frac{{\cal M} \ \eta}{5\ln10}\right)+ 7\ln\zeta(T_c) - \zeta(T_c)+\ln(T_c  \Delta t) \lsim 0
\label{key}
\ee
Note that equation (\ref{key}) is general and does
not depend on the precise form of the Friedmann equation. It only relies
on the adiabaticity of the expansion and on the radiation domination
assumptions. It is when replacing $\Delta t$ by
its expression in terms of the temperature that we make use of a
particular Friedmann equation.
Let us now express this condition using the standard Friedmann equation:
\be
H=\left(\frac{\dot a}{a}\right)=\sqrt{\frac{\rho(T)}{3 m_{Pl}^2}} 
\ee
$m_{Pl}=M_{Pl}/\sqrt{8\pi}$
 is the reduced Planck mass in four dimensions. This leads to 
\be
\rho(t)=\rho_0\left(1+\sqrt{\frac{4 \rho_0}{3m_{Pl}^2}}t\right)^{-2}
\approx \frac{3}{4}\frac{m_{Pl}^2}{t^2}
\label{rhooft}
\ee
where the last approximation applies if $\rho_0/\rho(t) \gg 1$ where $\rho_0$ is the energy 
density at some early time origin.
This gives us the relation between $t$ and $T$
\be
t=\frac{3}{2\pi}\sqrt{\frac{10}{g_*}}\frac{m_{Pl}}{T^2}\approx 0.14\frac{m_{Pl}}{T^2}
\label{tT}
\ee
and $\Delta t = \ t_c({T_c^2}/{T^2_2}-1)\sim 0.4 \ t_c  $.
Using (\ref{tT}) evaluated at $T_c$  and solving (\ref{key}) 
numerically with $\kappa=1/2,\ \alpha_w=0.0336, \ B=1.94, \  T_c=130$ GeV, we get the sphaleron bound:
\be
\zeta(T_c)\gsim 45.5 \ \ \rightarrow \ \ \frac{\phi(T_c)}{T_c}\gsim 1.21
\label{sphaleron bound}
\ee
where the second inequality was obtained using the fact that
$\zeta(T_c)$ is related to the order parameter $\phi(T_c)$ of the phase transition by:
\be
\zeta(T_c)=\sqrt{\frac{4\pi}{\alpha_w}}B\frac{\phi(T_c)}{T_c}
\ee
We can now check that with a more accurate evaluation of the integral in (\ref{integration}),
 equation (\ref{key}) is modified into\footnote{
$
I=\int^t_{t_c}V_B(t')dt'=2{\cal M}\eta \ \frac{0.14
 \ m_{pl}}{E_{sph}}\int^{\infty}_{\zeta_c}d\zeta\zeta^7e^{-\zeta}
$
where we treated $\phi(T)$ as a constant ($\phi(T)$ changes 
slowly with $T$), such approximation overestimates slightly the wash out. 
We noted $\zeta_c=\zeta(T_c)$. With
$
\int^{\infty}_{\zeta_c}d\zeta\zeta^7e^{-\zeta}=
\sum_{n=0}^7\frac{7 !}{n !}\zeta_c^ne^{-\zeta_c}\approx 
\zeta_c^7e^{-\zeta_c} \ \ \ (\zeta_c\gsim10)
$, we get $I=2 {\cal M} \eta t_c T_c \zeta_c^6e^{-\zeta_c}={2V_B(t_c)t_c}/{\zeta_c}$.
}
\be
\ln\left(\frac{2{\cal M} \ \eta}{5\ln10}\right)+ 6\ln\zeta(T_c) - \zeta(T_c)+\ln(T_ct_c) \lsim 0
\label{moreacurate}
\ee
weakening the sphaleron bound slightly: $\zeta_c\gsim 42.9 \ \rightarrow \ \phi_c/T_c\gsim 1.14$. 

\subsection{Particle physics constraints from the sphaleron bound.}

As stated in the introduction, the relation (\ref{sphaleron bound}) is usually translated into 
a bound on the higgs mass. Indeed, 
 ${\phi(T_c)}/{T_c}$ can be computed precisely from an analysis of the effective
potential of the higgs at high temperature. For instance, a simple perturbative evaluation at 
one loop gives
(with no {\it daisy} resummation) in the Standard Model \cite{Quiros:1999jp}
\be
\left(\frac{\phi(T_c)}{T_c}\right)_{SM}\approx\frac{2 (2m_W^3+m_Z^3)}{3m_h^2\times \pi v}
\ee
Plugging this expression into (\ref{sphaleron bound}) leads to $m_h \lsim 37$ GeV.
 As mentionned in the introduction, more precise calculations show 
that in the standard Model \cite{Kajantie:1996mn} there is no first 
order phase transition at all for higgs masses $m_h\gsim 72$ GeV. To make the phase 
transition first order, new bosonic
degrees of freedom with large couplings to the higgs are needed. The MSSM 
precisely provides these, in particular, the 
right stop $\tilde{t}_R$ with large Yukawa coupling can significantly enhance 
the strength of the phase transition\cite{Carena:1996wj,Carena:1997ki,Cline:1998hy}
}. Next formula (\ref{mssmstrength}) is 
an expression for 
${\phi(T_c)}/{T_c}$; it is not rigorously correct because it
 corresponds to a one-loop calculation and also to the ideal special case where 
 $m^2_U+c_sT^2\approx 0$ which means that 
the high temperature correction to the right stop mass $c_sT^2$ is compensated
 by its negative soft mass parameter $m^2_U$. However, it is helpful to get
 a flavor of what controls the strength of the EWPT in the MSSM:
 \be
\left(\frac{\phi(T_c)}{T_c}\right)_{MSSM}\sim\left(\frac{\phi(T_c)}{T_c}\right)_{SM} + 
\underbrace{\frac{2m_t^3}{\pi v
m_h^2}\left(1-\frac{\tilde{A}_t^2}{m_Q^2}\right)^{3/2}}_{\sim
\left[\frac{118 GeV}{m_h}\right]^2 \; \; \; \mbox{if} \; \; \; |\tilde{A}_t| \ll m_Q}
\label{mssmstrength}
\ee
Even if this one-loop evaluation underestimates ${\phi(T_c)}/{T_c}$ compared 
with lattice calculation we see from this expression that the condition of preservation
 of the baryon asymmetry sets serious constraints on $m_h$, $\tilde{A}_t/m_Q$ ($m_Q$ is the soft mass parameter of the left stop $\tilde{t}_L$) as well as 
$m_U^2$ (and therefore $m_{\tilde{t}_R}$). It is also instructive to keep in mind 
the one-loop expression for the Higgs mass (in the usual limit where the mass $m_A$ of the CP-odd neutral Higgs boson is large):
\be
m_h^2 = M_Z^2 \cos^2 2\beta + \frac{3}{4\pi^2}
\frac{\overline{m}_t^4}{v^2}
\log\left(\frac{m_{\widetilde{t}_R}^2 m_{\widetilde{t}_L}^2}
{\overline{m}_t^4}\right)\left[1
+ {\cal{O}}\left(\frac{\widetilde{A}_t^2}{m_Q^2}\right)
\right]
\label{susyhiggs}
\ee
While small values of $\tilde{A}_t/m_Q\lsim250/1000$
are required for baryogenesis, large values of $\tilde{A}_t/m_Q\gsim250/1000$ are required to 
satisfy the experimental bound on the higgs mass\footnote{An ultimate alternative 
to both satisfy the bound on the higgs mass and be consistent with small values 
of $\tilde{A}_t /m_Q$ would be to push the left stop 
mass to 2 or 3 TeV instead of 1 TeV.} and are in conflict 
with the sphaleron bound.

 The only alternative to reopen the
 window for electroweak baryogenesis is to weaken the sphaleron bound by modifying 
the last term in the left hand side of inequality (\ref{moreacurate}). For instance, to allow higgs masses 
in the range
$115 \; \mbox{GeV} \lsim m_h \lsim 118$ {GeV} with large
$ \tilde{A}_t /m_Q \sim 700/1000 $ and unconstrained right stop mass (\ie positive 
$ m_U^2$ values) 
one typically gets ${\phi(T_c)}/{T_c}\sim 0.3$ corresponding to  $\zeta(T_c)\sim 11.3$. 
Therefore
 one would need to weaken the sphaleron bound by a factor 4.

\subsection{Evading the sphaleron bound.}

Let us find out what value for the expansion rate would be needed at a 
temperature $T_c\sim 130$ GeV
to preserve the baryon asymmetry and be consistent with the 
larger window for the higgs and stop masses as presented below \ie when using 
 the value $\zeta(T_c)\sim 11$. Replacing $\zeta(T_c)\sim 11$ into (\ref{key}) leads to
\be
\ln\left(T_c \Delta t \right) \lsim 10.1
\label{bound}
\ee
In a radiation dominated era with the relation (\ref{tT}) we get 
\be
\ln (\frac{0.4T}{2H}) \lsim 10.1 \ \rightarrow \ H\gsim  10^{-3} \mbox{GeV}
\ee
while we have $H \sim 3 \ 10^{-14}$ GeV so $H$ is too small 
by 11 orders of magnitude at the EWPT.

Earlier proposals to weaken the sphaleron bound have been to relax the standard assumption 
of radiation domination at the EWPT and assume instead a universe dominated 
by the energy density of a scalar field \cite{Joyce:1996cp,Joyce:1997fc,Davidson:2000dw}.
The main observational constraint is that such domination must terminate by the 
nucleosynthesis epoch.
In  \cite{Joyce:1996cp,Joyce:1997fc}, it was shown that the energy density in a kinetic
 mode of a scalar field (which scales like $a(t)^{-6}$, faster than in the
 radiation case $a(t)^{-4}$) can significantly weaken the sphaleron bound and
 that such a phase of ``kinetion'' is required to evade large entropy release 
diluting the asymmetry. In \cite{Davidson:2000dw} it was argued that a universe 
dominated by the energy density of a more conventional scalar 
field (not necessarily in a kinetic mode) could also evade the erasure condition if the 
reheat temperature was as low as $T_r\sim $ MeV. However, the baryogenesis mechanism would have
 to be much more efficient than in the standard case thus standard electroweak baryogenesis does 
 not work in models with low reheat temperature and one would need to implement 
an Affleck--Dine mechanism for instance.

In this paper, we do not get rid of the radiation 
domination assumption and therefore do not have to assume any particular 
behaviour of a scalar 
field driving the expansion of the universe before nucleosynthesis.
Our point of view is quite different. We are instead allowing deviations to 
the Friedmann equation itself at early times
and study its consequences for electroweak baryogenesis.
Let us assume for example that before nucleosynthesis the expansion rate behaved like
\be
H\sim\frac{\rho(T)}{M^3}
\label{attempt}
\ee
where $M$ is some scale to be discussed later. Such
 behaviour may  {\it a priori} seem {\it ad hoc} 
but will be motivated in the next section. For a radiation dominated universe we get
\be
H\sim \frac{\pi^2g_*}{30}\frac{T^4}{M^3}
\ee
leading to
\be
\rho(t)\sim\rho_0\left(1+\frac{4  \ \rho_0 \ t}{M^3}\right)^{-1}
\label{rhooft2}
\ee
and the relation between time and temperature is now
\be
\label{Ttns}
t\sim\left(1-\left(\frac{g_*T}{g_*(T_0)T_0}\right)^4\right) \frac{15}{2\pi^2g_*}\frac{M^3}{T^4}
\approx\frac{15}{2\pi^2g_*}\frac{M^3}{T^4}
\ee
so that $\Delta t=t_c\left(\frac{T_c^4}{T_2^4}-1\right)\sim 0.95 \ t_c$. Equation (\ref{bound}) 
now translates into $M\lsim 20 $ TeV. Moreover, for $M\lsim 15$ TeV, 
the condition (\ref{key}) for the preservation of the baryon asymmetry is always 
satisfied whatever the value of $\zeta_c$\footnote{Weaker bounds on $M$ are obtained when
 evaluating more properly the integral (\ref{integration}) in the same way as 
equation (\ref{moreacurate}) was obtained.}. Before concluding that 
the expansion rate (\ref{attempt}) can successfully lower the sphaleron 
bound or even lead to the disappearance of that bound for sufficiently low scale $M$ we have
 to check two more things:
1) that sphalerons are at equilibrium in the symmetric phase \ie that baryon number 
violation is still completly efficient outside the bubble, 2) that electroweak interactions 
are at equilibrium as well
so that usual techniques to compute the baryon asymmetry continue
 to apply even in such rapidly expanding universe. 2) is actually automatically satisfied 
when 1) is satisfied since the sphaleron processes are the slowliest ones, being suppressed
 by a high power of $\alpha_w$; so we have to compare the rate of baryon number violation 
in the symmetric phase $V_B(t)^{sym}$ with the expansion rate $H$ and require:
\be
\label{symmetriccondition}
{\cal M} \ \Gamma_s^{sym}>H \ T^3 
\ee
According to latest calculations \cite{Bodeker:1999zt}:
\be
\label{guy}
\Gamma_s^{sym}=(25.4\pm 2)\alpha_w^5 T^4
\ee
Using the standard expansion rate of the universe one obtains from (\ref{symmetriccondition})
the condition $T <7 . 10^{12}$ GeV for sphaleron to be at equilibrium in the symmetric phase. On 
the other hand, for $M\sim 20 $ TeV using the expansion rate (\ref{attempt}) we get 
$T < 180$ GeV. This temperature is now close but still above the critical temperature. It
is actually sensitive to the exact prefactor in (\ref{guy}) as well as the value of the numerical factor 
 in the right hand side of equation
(\ref{attempt}) (which we took ${\cal O}(1)$ in a first attempt). So a more accurate
 analysis will be needed to determine which
 precise values of $M$ and the prefactor in $H$ are consistent with
 sphalerons at equilibrium in the symmetric phase.

One might also worry that in a fastly expanding universe it becomes difficult to
 nucleate bubbles and that the nucleation temperature $T_n$ might be quite 
different from the critical temperature $T_c$ at which the effective higgs 
potential has two degenerate minima. However, it has been shown \cite{Joyce:1997fc}
that the bubble nucleation temperature depends only weakly on the expansion rate 
of the universe so that $T_c\sim T_n$ remains valid even when varying the expansion 
rate $H$ over orders of magnitude. Consequently, the velocity of the phase 
interface may be subject to minor changes only. The same remark applies for
 other quantities which 
determine the baryon asymmetry generated like the wall thickness and perturbations in 
thermal population densities of particles at the vicinity of the wall.

From this rough analysis we remark that one can substantially weaken the sphaleron
 bound using 
the expansion rate (\ref{attempt}) as far as the scale $M$ is low enough.
Let us now study in which context this kind of behaviour might arise and how 
much freedom we have with respect to observational constraints.

\section{Two motivations to study deviations from Friedmann equation}

\noindent \underline{\it 1) No probe of FRW cosmology before nucleosynthesis}

\vspace{.5cm}
Friedmann's equation relates the expansion rate $H$ of the universe with 
its energy density $\rho(T)$ at a given temperature $T$. It 
is derived from the $00$ component of Einstein's equations in four dimensions, 
applied to the Robertson--Walker metric 
\be
ds^2=-dt^2+a^2(t)(\frac{dr^2}{1-kr^2}+r^2d\Omega^2)
\label{ansatz}
\ee
where $a(t)$ is the scale factor and $k$ the 
curvature parameter\footnote{In the rest of the paper we will set $k=0$.} and reads:
\be
H^2=\left(\frac{\dot a}{a}\right)^2=\frac{\rho(T)}{3 m_{Pl}^2} -\frac{k}{a^2}
\label{Fried}
\ee
Equation (\ref{Fried}) is a key ingredient of the standard model 
of cosmology, it serves as a 
basis to compute 
freeze-out temperatures and the density of relic particles in the universe. One of
 its major test comes from the success of the theory of primordial nucleosynthesis.
 However, there is no observational test of the Friedmann equation at times earlier 
than the primordial nucleosynthesis epoch characterized by an energy density 
$\rho \sim 0.1 \ \mbox{MeV}^4$, and it is legitimate to question the extrapolation 
of FRW cosmological evolution before that epoch. Therefore, one could, from an 
empirical point of view,
 allow deviations to the
 expansion rate of the universe of the form:
\be
H^2=\frac{\rho(T)}{3 m_{Pl}^2}(1+\alpha(T)) 
\label{parametrization}
\ee
where $\alpha(T)$ is a monotonic increasing function of temperature with a constrained
value at nucleosynthesis, $\alpha(T=T_{nucl}) \lsim 1/10$\cite{Olive:1999ij,Carroll:2001bv}. 
The origin of these 
deviations may come from
 modifications to the Robertson--Walker ansatz (\ref{ansatz}) by for example
questioning the homogeneity and isotropy assumptions or it may come from allowing 
modifications to general relativity itself as suggested by string theory. The existence 
of extra dimensions may induce corrections to the standard four-dimensional
 gravity, which are undetectable at low energy but might have played a 
significant role in the 
early universe.

\vspace{.5cm}
\noindent \underline{\it 2) Deviations to Friedmann equation arise in
 higher dimensional theories of gravity}

\vspace{.5cm}
The parametrization (\ref{parametrization}) is actually motivated by results
 obtained in the context of brane cosmology. 
Any higher dimensional theory of gravity which reproduces 4d Einstein gravity is expected
 to reproduce Friedmann--Robertson--Walker cosmology at least at late times. One famous example
is the 5d Randall--Sundrum (RS) model \cite{Randall:1999ee} where the warped non factorizable
 geometry allows 4d
 Einstein gravity to be recovered on the 3-brane. For instance, the effective non-relativistic
gravitation force between two masses $m_1$ and $m_2$ on the brane 
 obeys the usual Newton's potential plus some corrections scaling like ${\cal O}(r^{-3})$:
\be
V(r)=V_{\mbox{\tiny{Newton}}}(r)\left(1+\frac{1}{k^2r^2}\right), \ \ \ \   
V_{\mbox{\tiny{Newton}}}(r)=\frac{m_1m_2}{m_{Pl}^2}
\frac{1}{r}
\label{newton_RS}
\ee
where $k$ is the inverse of the AdS radius of the RS geometry. The validity of the Newton's 
formula has been checked down to distances $r\lsim  200 \ \mu m$ so that sets the 
constraint 
\be
\label{Newtonconstraint}
k\gsim 10^{-14} \mbox{GeV}
\ee
 The cosmology of the RS model has been studied by perturbing the
 static solution of the 5d Einstein equation and adding some energy density on the brane
responsible for the expansion\cite{Cline:1999ts,Csaki:1999jh}.
 It was shown that the expansion rate of the universe 
is of the form 
(\ref{parametrization}) with 
\be
\alpha(T)={\cal O}\left(\frac{\rho(T) m_{Pl}^2}{ M^6}\right)
\label{branecosmo}
\ee
where $M$ is the five dimensional Planck mass. 
It is remarkable that such a modified Friedmann equation
is not in conflict with primordial nucleosynthesis provided that 
$M\gsim 10$ TeV while it leads to drastic changes at early times 
for which $\alpha(T)$ dominates and therefore $H^2 \propto \rho^2$ instead of the 
conventional behaviour $H^2 \propto \rho$ \cite{Binetruy:1999ut,Chung:1999zs}. 
It turns out that a more severe constraint on $M$ comes from (\ref{Newtonconstraint})
which can be translated into $M\gsim 4 \ 10^4$ TeV by using the relation between the 5d and 4d Planck 
mass in the RS model $M^{3}=k M_{Pl}^2$. Such bound on $M$ is in conflict with our aim to weaken
 the sphaleron bound\footnote{Also, if one wants to use the RS geometry 
to solve the 
hierarchy problem then two branes are needed and $M\sim m_{Pl}$ which is much too large 
to evade the sphaleron bound.}. However, the result (\ref{parametrization}) with (\ref{branecosmo})
 can be generalized to any codimension 
one brane universe (for instance with vanishing 
tension and vanishing cosmological constant in the bulk \ie for a relatively small warping)
with compact stabilized fifth dimension. It has been extensively discussed in \cite{Csaki:1999mp}
that provided the $55$ component of Einstein equation has been used to stabilize the radion one
 generically recovers FRW cosmology for any codimension one brane model\footnote{In
 brane-world geometry 
one can find solutions to Einstein's equations where our four common 
dimensions only are expanding while the fifth extra dimension can be stabilized. 
This is to be contrasted with Kaluza--Klein cosmology where the energy density
 is uniformly distributed in all dimensions and leads to the expansion of the 
compact fifth dimension as well, consequently to a problematic time-dependant 
Newton `constant'.}. Let us consider the particular case of an Arkani-Hamed--Dimopoulos--Dvali 
type of geometry\cite{Arkani-Hamed:1998rs}. As is well-known, in the case with one extra dimension 
the fundamental Planck scale $M$ cannot be lowered to the TeV scale. This would lead to major 
modifications of gravity at observable distances. Therefore, if we want $M\sim 10-50$ TeV as 
required to lower
the sphaleron bound we need at least more than one large extra dimension.
So our point is to extrapolate the behaviour (\ref{branecosmo}) to brane models with a 
higher number $n$ of extra 
dimensions where $M$ can be lowered to a scale of order ${\cal O}$(10 TeV).
We do not know how the Friedmann equation looks like in the general case with $n>1$.
 While brane cosmology has been the subject of intense investigation in
 the last two years it is still at its infancy. Most studies have focused on the 5d case and
 also were essentially concerned with the recovery of the FRW cosmology at low energy. An important 
point to stress is that the result (\ref{parametrization},\ref{branecosmo}) is obtained 
first once the radion is stabilized and second in a perturbative approach by linearizing
 Einstein equations with respect to the perturbation $\delta a(t,y)$ (in $a(t,y)=a(y)_{RS}
+\delta a(t,y)$) and $\rho$.
What we are interested in in this paper is not the recovery of FRW but precisely the regime where
this perturbative expansion breaks down and the nature of the modification to the usual Friedmann 
equation. A specific feature of brane cosmology in 5d are the $\rho^2$ term corrections. They are
 somehow a consequence of the junction conditions on the brane\cite{Binetruy:1999ut} which are 
derived by 
identifying the singularity of the second derivative of the metric $a''/a$ with the singularity of the
 energy-momentum tensor. To determine the junction conditions in the higher dimensional case one would
need to solve equations of the type $\Delta a(y_1,...,y_n))/a \sim \rho \delta^n(y_1,...,y_n)$ where
 $\Delta$ is the
 laplacian in $n$ dimensions. Being ignorant about the non static solutions of Einstein equations 
in the $n>1$ case we will assume 
that $\rho^2$ terms still arise in Friedmann equation and are suppressed by $M^6$ as in the 5d case 
so that they dominate over the usual term at early times and in particular at the time of the EWPT. 
We will therefore use the expression
\be
H=\sigma \ \frac{\rho}{M^3}
\label{Hns}
\ee
where $\sigma$ parametrizes our ignorance on the exact numerical factor; we will focus on a typical 
range of values $[0.1-1]$.

\section{Some comments on the cosmology with non standard Friedmann equation}

Note that the transition between the standard and non standard cosmologies would take place when 
${\sqrt{\rho}}/{\sqrt{3}m_{Pl}}\sim{\sigma \rho}/{M^3}$.
In a radiation dominated era this leads to
\be
T_{\mbox{\tiny{trans}}}\sim \left(\frac{M^3}{m_{Pl}}\frac{\sqrt{10}}
{\pi\sigma\sqrt{g_*}} \right)^{1/2}
\ee
With $\sigma=1/5$, $M=20 $ TeV we get $T_{\mbox{\tiny{trans}}} \sim 1.4$ MeV 
\ie $\rho\sim 135$ MeV$^4\sim1350 \ \rho_{\mbox{\tiny{nucl}}}$. Therefore, transition to conventional 
cosmology takes place slightly before nucleosynthesis.

While it is reasonable to expect that the highest temperature in the universe 
$T_{\mbox{\tiny{reheat}}}$ was ${\cal O}(M)$ it has been argued that in theories with low
 fundamental Planck scale $T_{\mbox{\tiny{reheat}}}$ has to be much lower ($\lsim 1$ GeV)
to be compatible with observational constraints\cite{Arkani-Hamed:1998nn,Benakli:1998ur}. This is
 due to the presence of light modes other 
than standard model particles which can propagate in extra dimensions, in particular the graviton, 
and can affect substantially the energy density of the radiation after the decay of the inflaton 
on our brane. There are different independant kinds of bounds on $T_{\mbox{\tiny{reheat}}}$. They come 
from requiring that 1) the energy density of Kaluza--Klein gravitons does not modify the 
expansion rate of the universe at the time of nucleosynthesis, 2) the decay of KK gravitons into
 photons does not generate distorsions in the $2.7 K$ photon background radiation, 3) the energy density 
of KK gravitons does not lead to overclosure of the universe, 4) the cooling of the energy density 
of the radiation on our brane is dominated by the expansion and not the escape of gravitons into the
 bulk. 
All these constraints are based on the assumption that the bulk is empty (and static) and 
gravitons can only decay on our brane. Therefore they can be evaded if assuming that graviton can 
decay on other brane than ours. Also, a toroidal geometry is assumed for extra dimensions. Other 
choices of geometry lead to different spectra of KK modes which lead to weaker
bounds on $T_{\mbox{\tiny{reheat}}}$ \cite{Kaloper:2000jb} and the occurrence of the 
EWPT in a radiation dominated era is no more questionned\footnote{For other discussions on the reheating 
temperature in non conventional brane cosmologies see also \cite{Allahverdi:2001eq}.}.

It is interesting however to come back on condition 4). In our previous analysis of subsection 2.2, 
the relation between time and temperature was derived assuming that cooling is due to 
the expansion of the universe so that $\rho(t)\sim a(t)^{-4}$ and
$\dot{\rho}|_{\mbox{\tiny{exp}}}=-4H\rho$. The cooling due to the escape of gravitons into the
 extra dimensions was evaluated in \cite{Arkani-Hamed:1998nn}:
\be
 \dot{\rho}|_{\mbox{\tiny{esc}}}\sim-T^{n+7}/M^{n+2}
\label{rough}
\ee
 Requiring that the cooling by escape of 
graviton is less than ten percents leads to 
$H\rho\gsim 5T^{n+7}/2M^{n+2}$. For a standard cosmology this gives 
\be
T^{n+1}\lsim \frac{2}{5\sqrt{3}}\left(\frac{\pi^2 g_*}{30}\right)^{3/2}\frac{M^{n+2}}{m_{Pl}}
\ee
If $M=10$ TeV we obtain $T \lsim 13$ GeV for $n=2$. On the 
other hand using the expansion rate (\ref{Hns}) we get
\be
T^{n-1}\lsim \frac{2}{5}\sigma\left(\frac{\pi^2g_*}{30}\right)^2 M^{n-1}
\ee
So, even if we were to assume the validity of (\ref{rough}), at any temperature below $M$, the
 cooling is well dominated by the expansion.

Let us finish this section with the evaluation on the lower bound on $M$ obtained by 
requiring that 
the expansion 
rate (\ref{Hns}) should be suppressed at the time of nucleosynthesis when $\rho\sim 0.1$ MeV$^4$. 
Requiring $\alpha(T)=3\sigma^2\rho m^2_{Pl}/M^6\lsim 1/10$ leads to
\be
M\gsim \sigma^{1/3} \ 16.1 \ \mbox{TeV}
\label{nucleobound}
\ee
so that  for $\sigma=1$,  $M\gsim 16.1$ TeV,  for $\sigma=1/5$,  $M\gsim 9.4$ TeV and for $\sigma=1/10$, 
 $M\gsim 7.4$ TeV. 

In the next section we refine the rough analysis of subsection 2.2. Our aim is to study the
 parameter space leading to a significant weakening of the sphaleron bound. There are 
three parameters to work with: The higher dimensional Planck mass $M$, the factor $\sigma$ 
appearing in the expansion rate (\ref{Hns}) and $\kappa$ the prefactor in the 
sphaleron rate (\ref{sph_rate}). We will limit our study to values of 
$\kappa$ in the range $[0.1-1]$ since non perturbative calculations of the sphaleron 
rate indicate $\kappa\gsim 0.1$.

\section{Condition to preserve the baryon asymmetry using non conventional cosmology}

\begin{figure}[h]
\rotatebox{270}{\includegraphics[height=18cm]{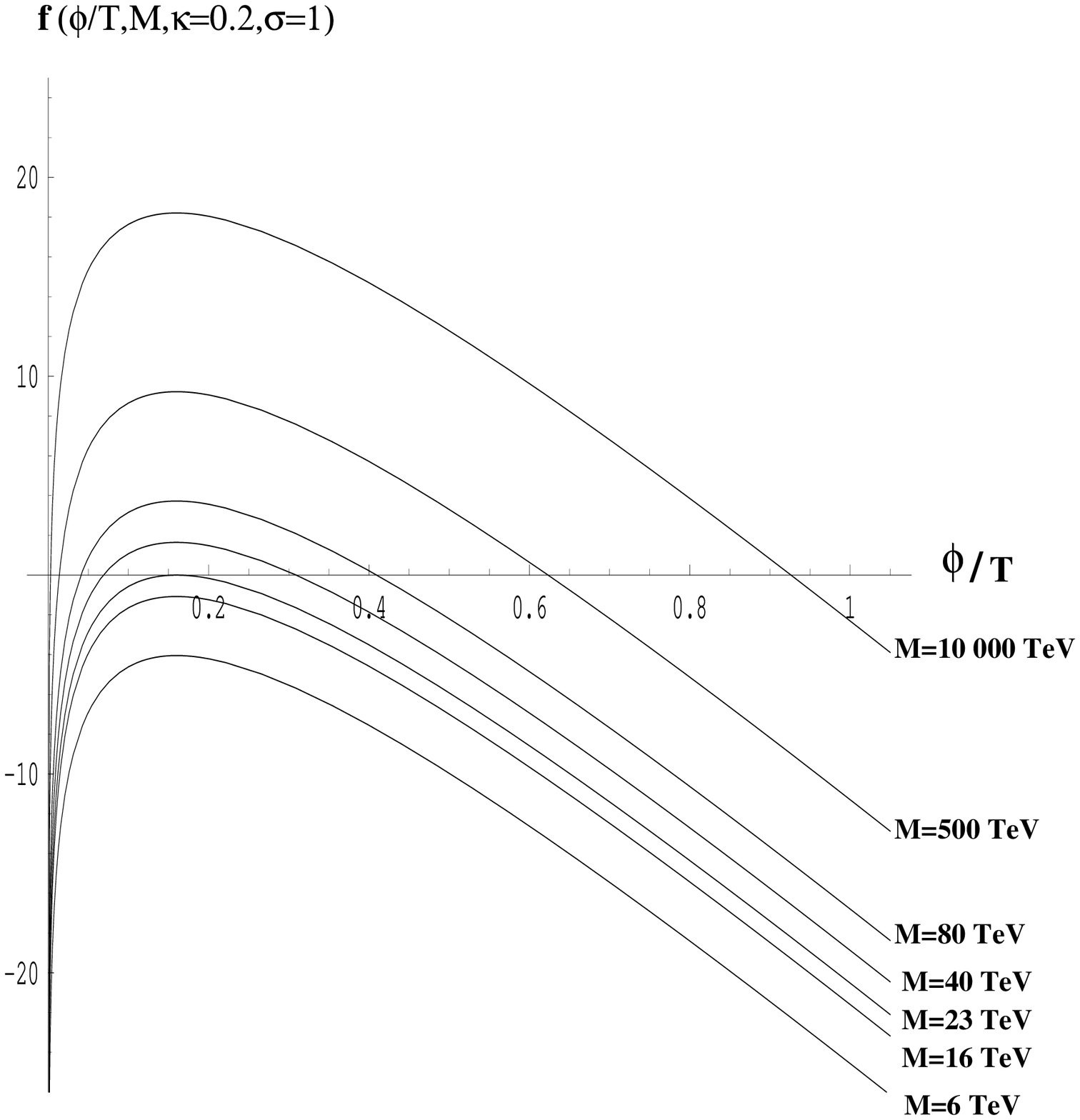}}

\caption[]{$f(\frac{\phi_c}{T_c},M,\kappa,\sigma)$ expresses the condition of preservation of the 
baryon asymmetry. For negative values of $f$, baryon asymmetry is preserved. This leads to a bound on 
the value of $\phi_c/T_c$ for $M\gsim23$ TeV which depends on the fundamental Planck scale. As stated at
the end of section 5, one should not pay attention to what happens to the graph 
for $\phi_c/T_c\lsim 0.15$. The turnover is an artifact of the breakdown of 
the approximation leading to formula (\ref{gammas}).}
\label{fig:f(phi/T)}
\end{figure}

Let us now analyse in more details the condition ${\cal S} \gsim 10^{-5}$ expressing the preservation 
of the baryon asymmetry produced at the EWPT. The integral in (\ref{integration}) is now
 $I\approx 4{\cal M}\eta T_c t_c \zeta_c^6e^{-\zeta_c}$, leading to
\be
\ln\left(\frac{4{\cal M} \ \eta}{5\ln10}\right)+ 6\ln\zeta(T_c) - \zeta(T_c)+\ln(T_ct_c) \lsim 0
\label{moreacurate2}
\ee
which is very similar to (\ref{moreacurate}). It can be rewritten as
\be
f(\frac{\phi_c}{T_c},M,\kappa,\sigma)\lsim 0
\ee
where
\be f(\frac{\phi_c}{T_c},M,\kappa,\sigma)=
3.06+\ln(\frac{\kappa \ M^3}{\sigma T^3_c})+6\ln(\frac{\phi_c}{T_c})-37.52 \ \frac{\phi_c}{T_c}
\ee
This last expression has been obtained by replacing $t_c$
in (\ref{moreacurate2}) 
by its expression (\ref{Ttns}) and that is what makes the difference 
with the conventional cosmology.
In figure \ref{fig:f(phi/T)} we plotted, for different values of $M$, 
$f(\frac{\phi_c}{T_c},M,\kappa,\sigma)$ as a function of $\frac{\phi_c}{T_c}$
in the case $\sigma=1$, $\kappa=0.2$. We find that for
 $M\lsim 23.1$ TeV, $f$ is always negative which means that the condition of preservation of the 
baryon asymmetry is always satisfied whatever the value for $\phi_c/T_c$. Therefore, 
  if $\sigma=1$ and $M$ is in the range $[16.1-23.1]$ TeV there
 is no sphaleron bound.
For larger values of $M$, $f$ becomes positive
 for certain values of $\phi_c/T_c$ and therefore there is a sphaleron bound. Notice 
that $M=23.1$ TeV corresponds to a discontinuity in the bound on ${\phi_c}/{T_c}$. From no constraint on 
${\phi_c}/{T_c}$ we jump to the constraint $\phi_c/T_c>0.16$.

\begin{figure}[h]
\centering
\includegraphics[height=10cm,width=9.5cm]{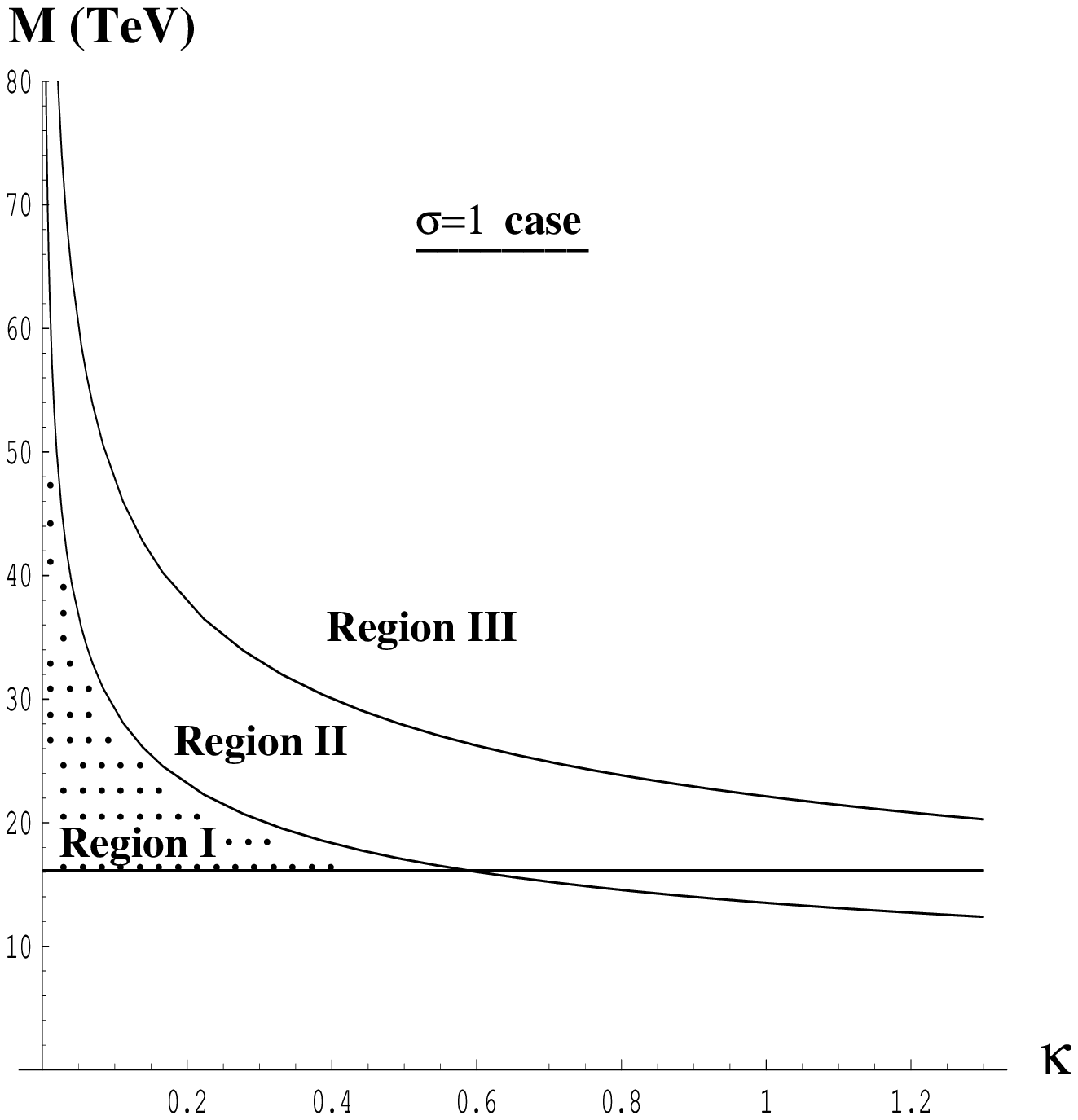}
\includegraphics[height=10cm,width=9.5cm]{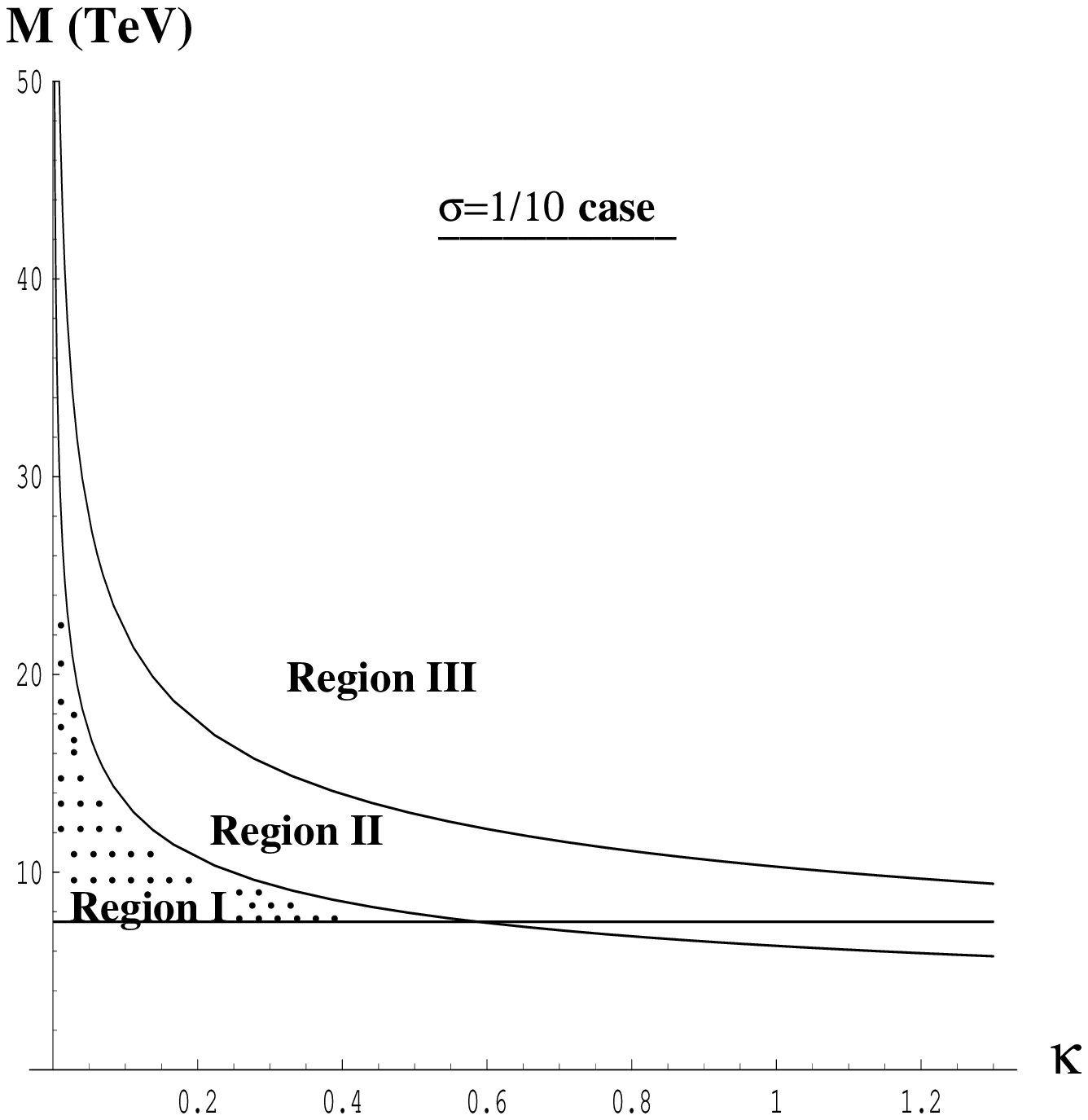}
\caption[]{Sphaleron bound in the parameter space $(\kappa,M)$. In Region I, there 
is no condition at all on $\phi_c/T_c$. 
In region II, $0.16\le\phi_c/T_c\le 0.3$. In region III, 
$\phi_c/T_c>0.3$.
}
\label{fig:parameterspace}
\end{figure}

In figure \ref{fig:parameterspace}, we summarize the status of the sphaleron bound in different
regions of the parameter space $(M,\kappa)$ for the two cases $\sigma=1$ and $\sigma=1/10$. The 
horizontal line corresponds to the lower bound on
 $M$ dictated by compatibility with nucleosynthesis. Another type of lower bound on $M$ is obtained
 by requiring that sphalerons are at thermal
 equilibrium in the symmetric phase (see equation (\ref{symmetriccondition}))
at $T_c$. It leads to the condition
$
M>0.1 \ T_c \ \sigma^{1/3}
$. For $T_c=130$ GeV we get $
M>14.3\ \sigma^{1/3}
$TeV, which is slightly weaker than the bound obtained from nucleosynthesis. Anyhow, it is 
interesting to point out that independently from the nucleosynthesis constraint, what electroweak
 baryogenesis singles out is a window of a few tens of TeV for $M$. The upper bound corresponds 
to an expansion rate which is too small to freeze the baryon asymmetry in the broken phase at the
 EWPT while the lower bound corresponds to an expansion rate which is so large that sphalerons
 become out of equilibrium in the symmetric phase as well.
The upper curves corresponds to $\phi_c/T_c=0.3$ 
which we presented in section 2 as the value obtained for higgs masses as large as 115 GeV and 
stop masses no more constrained to be lighter than the top mass. The lower curve delimits the regions
 where there is a sphaleron bound at least $\phi_c / T_c\gsim 0.15$, and no sphaleron bound at all.

We should add a last clarifying remark here. The main formula we have made use of in our analysis is equation
(\ref{gammas}) which gives the rate of baryon number violation in the broken phase. One should keep in
 mind that this formula is valid in a relatively narrow temperature interval $M_W(T)<<T<<E_{sp}(T)$. As
 stated in \cite{Arnold:1987mh,Khlebnikov:sr}, for $E_{sp}/T\lsim1$ \ie for $\phi/T\lsim0.1$, configurations 
other than the sphaleron 
contribute to the probability of making a transition with changing baryon number. Consequently, we
 can no longer apply the analysis of decay of metastable states \cite{Langer:bc} with allowance for only the 
saddle point (\ie sphaleron solution) of the energy functional. As $\phi/T$ reaches zero, the 
sphaleron rate is expected to approach its value in the symmetric phase and the turnover 
around $\phi/T \sim$ 0.15 in the graph of figure 1 looses meaning. This however does not alter the validity 
of our qualitative results. Moreover, in the range $\phi/T \sim 0.2-0.4$, formula (\ref{gammas}) leads 
to an overestimation of the value of 
the sphaleron rate (it slightly overpasses the
 value in the symmetric phase) which means that by using that formula in that regime 
we have slightly underestimated the freeze out.

\section{Conclusion}

In summary, we studied the condition for preservation of the baryon asymmetry produced 
at the EWPT when questioning the usual assumption on the thermal
 history of the universe. To weaken the sphaleron bound by a factor 3 or 4 as required to 
allow higgs masses as large as 115 GeV as well as a wider window of 
the MSSM parameter space, the expansion rate of the universe at the EWPT has
 to be increased dramatically (by ten orders of magnitude at least) in order to freeze 
out the sphaleron processes in the $SU(2)$ broken phase. On the other hand, the expansion
 rate should not be larger than a critical value above which sphalerons start to be 
inoperative in the symmetric phase as well. It turns out that the the expansion rate (\ref{Hns})
inspired by brane cosmology possesses these properties for $M$ in the approximate 
range $15-50$ TeV depending on the precise values of the parameters $\kappa $ and $\sigma$. It would
 be interesting to check that the Friedmann equation we have postulated effectively arise in
 generic brane models and more generally to study 
deviations to Friedmann equation before the nucleosynthesis epoch in the spirit of \cite{Carroll:2001bv}
which do not necessarily rely on the existence 
of extra dimensions.

The possibility we have investigated is one
among others. For instance, we are still assuming radiation domination. One could imagine various ways 
to modify the cosmology at the EWPT\footnote{One could also exploit the fact that for small
 values of $M$, sphalerons are frozen even in the symmetric phase and cannot erase
 any preexisting baryon asymmetry 
 produced 
by mechanisms which do not make use of a first order phase transition.}.
Also, in our work, $M$ is the fundamental Planck scale, but it could correspond to a 
different scale. For example, the authors of  \cite{Csaki:1999mp} found that the $\rho^2$ 
corrections 
in Friedmann equation could be suppressed by powers of $\Lambda_W$ (in a matter dominated 
era) where 
$\Lambda_W$ is a scale associated with the mass of the radion. Note that we are considering 
a regime where gravity becomes essentially higher dimensional and where radion modes could play 
a role. One thing to care about is that decay of these modes will not dilute the baryon asymmetry.

Finally, we have focused here on the ``out of equilibrium'' condition of electroweak baryogenesis.
There is still much activity in the electroweak baryogenesis community to determine whether
 there are enough CP violation sources in the MSSM to account for the amount of baryon 
asymmetry observed today ($n_B/s\sim 10^{-11}$).

\section*{Acknowledgements}
I am very grateful to Carlos Wagner for stimulating and instructive discussions. It is also 
a pleasure to thank Sean Carroll, Jim Cline, Christophe Grojean and Mariano Quiros 
for helpful comments.
This work is supported in part by the US Department of Energy, High Energy Physics
Division, under contract W-31-109-Eng-38 and also by the David and Lucile Packard
Foundation.


\end{document}